\documentclass{natureprintstyle}
\usepackage{epsfig,caption}
\usepackage{color}
\usepackage{bm}
\usepackage{graphicx}
\usepackage{longtable}
\usepackage{amssymb}
\usepackage{rotating}
\usepackage{latexsym}
\usepackage{hyperref}
\usepackage{float}

\newcommand{\rmaxs}{\ifmmode{R_{\rm{\sigma}}^{\rm{max}}}\else{$R_{\rm{\sigma}}^{\rm{max}}$}\fi}
\newcommand{\recirc}{\ifmmode{R_{\rm{e,c}}}\else{$R_{\rm{e,c}}$}\fi}
\newcommand{\re}{\ifmmode{R_{\rm{e}}}\else{$R_{\rm{e}}$}\fi}
\newcommand{\ee}{\ifmmode{\epsilon_{\rm{e}}}\else{$\epsilon_{\rm{e}}$}\fi}
\newcommand{\lr}{\ifmmode{\lambda_R}\else{$\lambda_{R}$}\fi}
\newcommand{\lre}{\ifmmode{\lambda_{R_{\rm{e}}}}\else{$\lambda_{R_{\rm{e}}}$}\fi}
\newcommand{\vs}{\ifmmode{V / \sigma}\else{$V / \sigma$}\fi}
\newcommand{\vse}{\ifmmode{(V / \sigma)_{\rm{e}}}\else{$(V / \sigma)_{\rm{e}}$}\fi}
\newcommand{\vobs}{\ifmmode{V_{\rm{obs}}}\else{$V_{\rm{obs}}$}\fi}
\newcommand{\sobs}{\ifmmode{\sigma_{\rm{obs}}}\else{$\sigma_{\rm{obs}}$}\fi}

\newcommand{\kms}{\ifmmode{\,\rm{km}\, \rm{s}^{-1}}\else{$\,$km$\,$s$^{-1}$}\fi}
\newcommand{\msun}{\ifmmode{M_{\odot}}\else{$M_{\odot}$}\fi}
\newcommand{\at}{\ifmmode{\rm{ATLAS}^{\rm{3D}}}\else{ATLAS$^{\rm{3D}}$}\fi}

\newcommand{\farcs}{\mbox{$.\!\!^{\prime\prime}$}}

\title{A relation between characteristic stellar age of galaxies and their intrinsic shape}

\author{Jesse van de Sande$^{1}$, 
Nicholas Scott$^{1,2}$, 
Joss Bland-Hawthorn$^{1}$, 
Sarah Brough$^{2,3},$
Julia J. Bryant$^{1,2,4},$
Matthew Colless$^{2,5}$,
Luca Cortese$^{6},$
Scott M. Croom$^{1,2},$
Francesco d'Eugenio$^{2,5},$
Caroline Foster$^{1,7},$
Michael Goodwin$^{4},$
Iraklis S. Konstantopoulos$^{4,8},$
Jon S. Lawrence$^{4},$
Richard M. McDermid$^{4,9},$
Anne M. Medling$^{5,10,11},$
Matt S. Owers$^{4,9},$ 
Samuel N. Richards$^{12},$
Rob Sharp$^{2,5}$
}

\begin{document}

\maketitle

\let\thefootnote\relax\footnote{
\begin{affiliations}
\item {Sydney Institute for Astronomy, School of Physics, A28, The University of Sydney, NSW, 2006, Australia} 
\item {ARC Centre of Excellence for All-Sky Astrophysics (CAASTRO)}
\item {School of Physics, University of New South Wales, NSW 2052, Australia}
\item {Australian Astronomical Observatory, PO Box 915, North Ryde NSW 1670, Australia}
\item {Research School of Astronomy and Astrophysics, Australian National University, Canberra ACT 2611, Australia}
\item {International Centre for Radio Astronomy Research, The University of Western Australia, 35 Stirling Highway, Crawley WA 6009, Australia}
\item {ARC Centre of Excellence for All Sky Astrophysics in 3 Dimensions (ASTRO 3D)}
\item {Atlassian 341 George St Sydney, NSW 2000}
\item {Department of Physics and Astronomy, Macquarie University, NSW 2109, Australia}
\item {Cahill Center for Astronomy and Astrophysics California Institute of Technology, MS 249-17 Pasadena, CA 91125, USA}
\item {Hubble Fellow}
\item {SOFIA Operations Center, USRA, NASA Armstrong Flight Research Center, 2825 East Avenue P, Palmdale, CA 93550, USA}
\end{affiliations}
}

\begin{abstract}
Stellar population and stellar kinematic studies provide unique but complementary insights into how galaxies build-up their stellar mass and angular momentum\cite{tinsley1980,davies1983,bender1993}. A galaxy's mean stellar age reveals when stars were formed, but provides little constraint on how the galaxy's mass was assembled. Resolved stellar dynamics\cite{cappellari2016} trace the change in angular momentum and orbital distribution of stars due to mergers, but major mergers tend to obscure the effect of earlier interactions\cite{naab2014}. With the rise of large multi-object integral field spectroscopic (IFS) surveys, such as SAMI\cite{croom2012} and MaNGA\cite{bundy2015}, and single-object IFS surveys (e.g., ATLAS$^{\rm{3D}}$ \cite{cappellari2011a}, CALIFA\cite{sanchez2012}, MASSIVE\cite{ma2014}), it is now feasible to connect a galaxy's star formation and merger history on the same resolved physical scales, over a large range in galaxy mass, and across the full range of optical morphology and environment\cite{cappellari2016,vandesande2017,scott2017}. Using the SAMI Galaxy Survey, here we present the first study of spatially-resolved stellar kinematics and global stellar populations in a large IFS galaxy survey. We find a strong correlation of stellar population age with location in the (\vs, \ee) diagram that links the ratio of ordered rotation to random motions in a galaxy to its observed ellipticity. For the large majority of galaxies that are oblate rotating spheroids, we find that characteristic stellar age follows the intrinsic ellipticity of galaxies remarkably well. This trend is still observed when galaxies are separated into early-type and late-type samples.

\end{abstract}

Using the SAMI Galaxy Survey IFS data, we measure the spatially resolved stellar rotation $V$ (\kms), stellar velocity dispersion $\sigma$ (\kms)\cite{vandesande2017}, and global stellar population luminosity-weighted age (Gyr)\cite{scott2017}. Stellar masses are derived from the relation between optical $g-i$ colour, $i$-band luminosity, and stellar mass. Our sample of 843 galaxies covers a broad range in stellar mass ($9.5<\log M_{*}/\msun<11.6$), and the full range in optical morphology (E-Sd), and large-scale environment (field-cluster).

\begin{figure}[!b]
\begin{center}
\includegraphics[width=\linewidth]{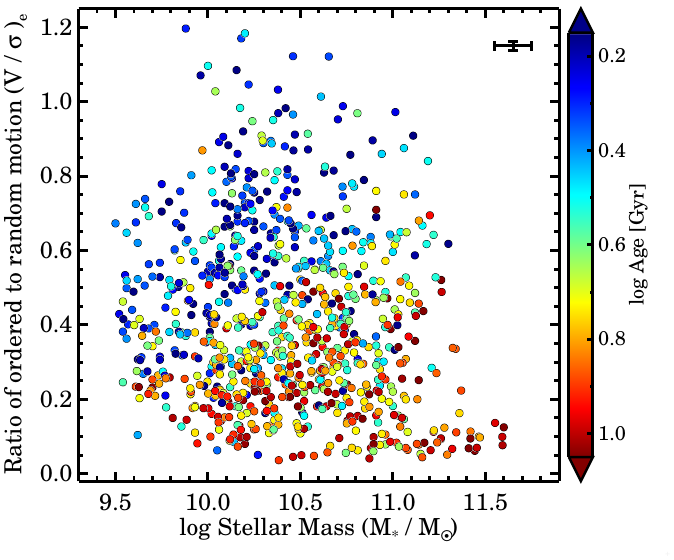} 
\caption{Ratio between stellar ordered rotation and random orbital motion in a galaxy \vse\ as a function of total stellar mass. Data are colour coded by the luminosity-weighted stellar population age within one effective radius. The median uncertainty on \vse\ and $M_*$ is shown in the top-right corner; the median uncertainty on $\log \rm{Age}$ is $\pm0.15$ dex.}
\label{fig:vs_lmass_age}
\end{center}
\end{figure}

In Figure \ref{fig:vs_lmass_age}, we present our stellar kinematic measurements \vse\ as a function of stellar mass. \vse\ quantifies the ratio between ordered rotation and random orbital motion in a galaxy and is defined as the square root of the ratio of the luminosity-weighted $V^2$ and $\sigma^2$ within an ellipse that encloses half of the projected total galaxy light (Eq.~\ref{eq:vs})\cite{binney2005,cappellari2007}. We colour code the data by the luminosity-weighted stellar age measured from an integrated spectrum within the same effective elliptical aperture for which the stellar kinematics are derived. There is a strong relation between \vse\, and age, such that galaxies with young stellar populations (Age$<2.5$ Gyr, blue) are predominantly rotationally supported (mean $\vse=0.58$), whereas galaxies with old stellar populations (Age$>10$ Gyr, red) are more pressure supported by random orbital motion of stars (mean $\vse=0.23$).

\begin{figure*}[!t]
\begin{center}
\includegraphics[width=\linewidth]{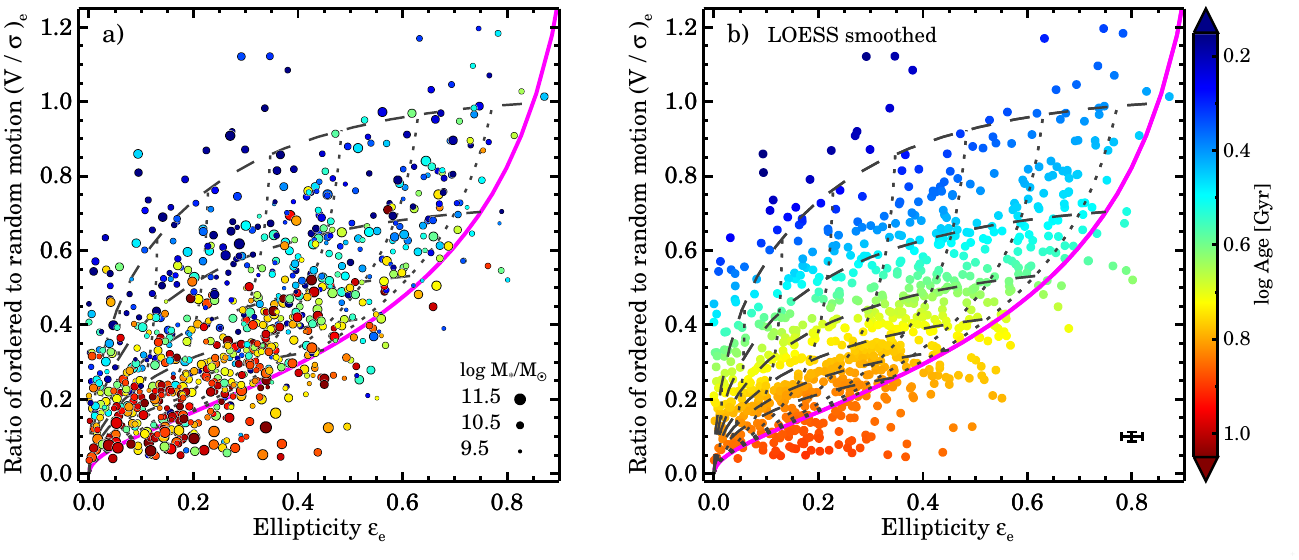} 
\caption{Linking stellar dynamics \vse\ and observed shape (ellipticity $\ee$) with luminosity-weighted stellar age within one effective radius. In panel a) the colour coding reflects young (blue) and old (red) stellar population age of each individual galaxy and symbol size indicates the total stellar mass of the galaxy, whereas in panel b) we use the LOESS smoothing algorithm to recover the mean underlying trend in age.
The median uncertainty on \vse\ and \ee\ is shown in the bottom-right corner of panel b), and the median uncertainty on $\log \rm{Age}$ is $\pm$0.15 dex. Theoretical predictions for the edge-on view of axisymmetric galaxies with anisotropy $\beta_z = 0.6\times \epsilon_{\rm{intr}}$ are shown as the solid magenta line\cite{cappellari2007}. The dotted lines show the model with different viewing angle from edge-on (magenta line) to face-on (towards zero ellipticity). Galaxies with different intrinsic ellipticities $\epsilon_{\rm{intr}}$=0.85-0.35 (top to bottom) are indicated by the dashed lines.
}
\label{fig:vs_eps_age}
\end{center}
\end{figure*}

\begin{figure*}[!t]
\begin{center}
\includegraphics[width=\linewidth]{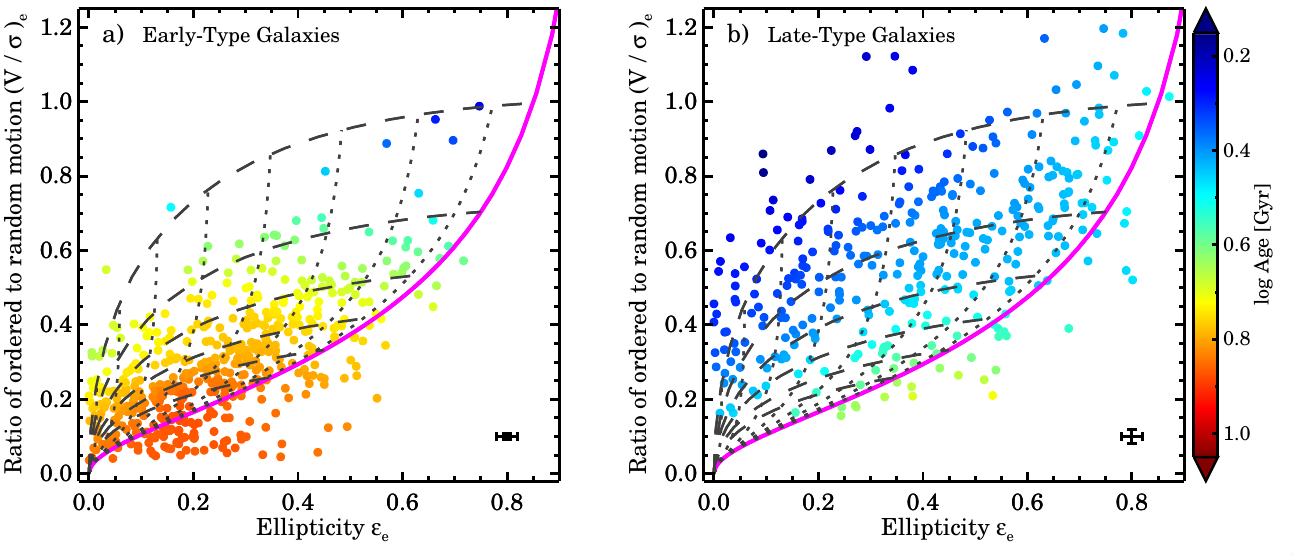} 
\caption{Luminosity-weighted stellar age in the (\vs, \ee) diagram, split by visual morphological type. Galaxies with early-type morphology (Ellipticals and S0s) are shown in panel a), galaxies with late-type morphology (Sa-Sd, irregulars) are shown in panel b). The colour coding highlights the LOESS smoothed stellar population ages. The LOESS algorithm is applied separately to the two individual subsamples. Lines similar to Figure \ref{fig:vs_eps_age}.
}
\label{fig:vs_eps_age_es}
\end{center}
\end{figure*}

While the relation between age and \vse\ is stronger than the relation between age and stellar mass (Supplementary Figure \ref{fig:sup_age_mass_vs_scatter}), both relations still leave a large residual. In Figure \ref{fig:vs_eps_age}, we investigate the connection between age and stellar dynamical properties further. We show the (\vs, \ee) diagram for galaxies that relates the ratio of ordered to random motion \vse\ versus the observed ellipticity \ee. Besides colour-coding galaxies by their individual ages in panel a), we now also use a locally weighted regression algorithm (LOESS\cite{cappellari2013b}) to recover the mean underlying trend in the sample as shown in panel b). Additionally, we show theoretical predictions from the tensor virial theorem that links velocity anisotropy, rotation and intrinsic shape\cite{binney2005}. We use the best-fitting relation from a high-quality subset of galaxies from the SAURON sample $\beta_z = 0.6\pm 0.1\times \epsilon_{\rm{intr}}$\cite{cappellari2007}, i.e., we assume that all galaxies are mildly anisotropic. An axisymmetric, oblate rotating spheroid with varying \textit{intrinsic} ellipticity and anisotropy, when observed edge-on, is shown as the solid magenta line\cite{cappellari2007}. On this line, nearly round spherical galaxies reside on the bottom left; flattened rotating disks are on the top right. The dashed lines show galaxies with constant intrinsic ellipticities, but observed with varying viewing angle from face-on (zero ellipticity) to edge-on (towards the magenta line).

LOESS smoothing reveals that the age of a stellar population follows the lines of different projected intrinsic ellipticities remarkably well; along a line of constant intrinsic ellipticity, smoothed age is nearly constant. As the shape and kinematic properties of the galaxies transform from dynamically cold, and intrinsically flat into dynamically-hotter, pressure-supported, thicker, oblate spheroids, the mean age of the stellar population is seen to increase. On the bottom left part of the diagram, we find maximally old, fully dispersion-dominated, nearly-spherical galaxies. 

We simulate the minimum required sample size for detecting this trend, by randomly drawing galaxies from our full sample of 843 galaxies, and then repeating the LOESS smoothing on that subsample. The relation between age and intrinsic ellipticity becomes visible with a minimum number of $\sim250$ galaxies, but only if the sample has a full range in visual morphology.

There is no dependency on stellar mass within the (\vs, \ee) diagram for all galaxies that are consistent with being axisymmetric, rotating, oblate spheroids (above magenta line), though galaxies with the highest stellar mass are more likely to be pressure-supported spheroids than rotating (Supplementary Figure \ref{fig:sup_vs_eps_age_massbins}-\ref{fig:sup_vs_eps_mass}).

\begin{figure*}[!ht]
\begin{center}
\includegraphics[width=\linewidth]{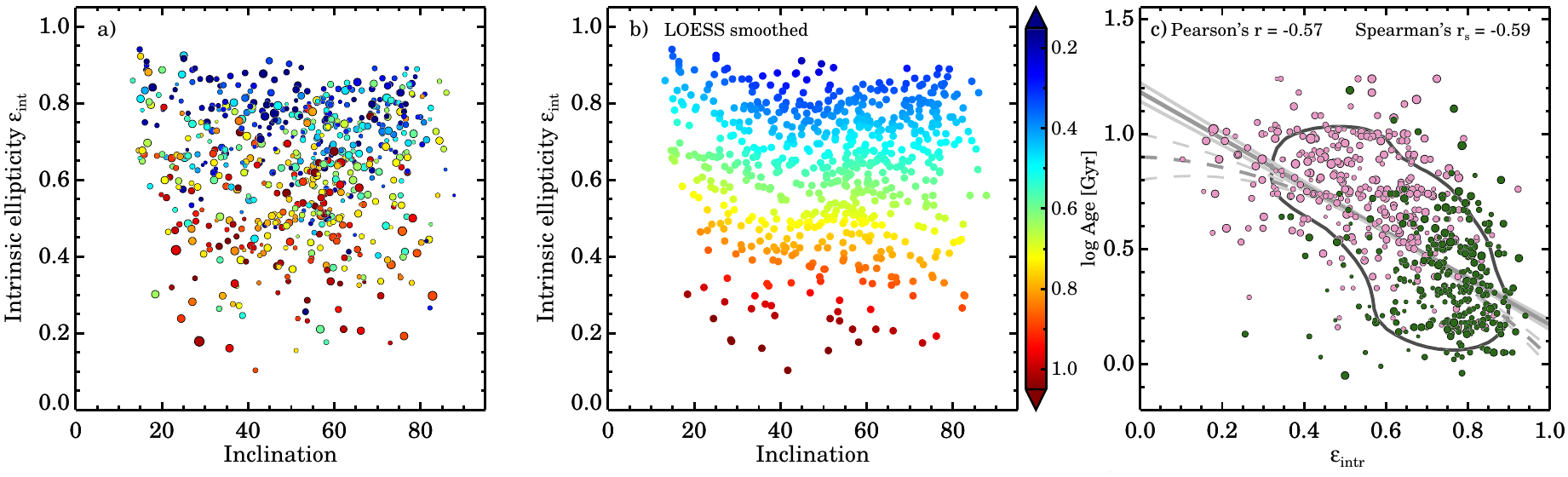} 
\caption{Intrinsic ellipticity and inclination derived from theoretical predictions and our data from within the (\vs, \ee) diagram. Only galaxies that are consistent with being axisymmetric rotating spheroids, and with $\ee>0.025$ are included in the sample. In panel a) the colour coding reflects the stellar population age of individual galaxies and symbol size indicates the total stellar mass of the galaxy, whereas in panel b) we use the LOESS smoothing algorithm to recover the mean underlying trend in age.
We show the relation between age and intrinsic ellipticity in panel c) for early-type (pink) and late-type (green) galaxies. The contour encloses 68\% of the total probability using kernel density estimates. The solid line shows the best-fitting relation of $\log \rm{Age} = -1.01\pm0.06\times\epsilon_{\rm{intr}} + 1.18\pm0.04$, with an RMS scatter in $\log \rm{Age}$ of 0.238 dex, using the \textit{Interactive Data Language} \textsc{LINMIX\_ERR} function that uses Bayesian approach to linear regression. The dashed line shows the best-fitting relation $\log \rm{Age} = -0.80\pm0.28\times\epsilon_{\rm{intr}}^2 - 0.07\pm0.34\times\epsilon_{\rm{intr}} + 0.90\pm0.10$ using the \textsc{POLY\_FIT} function, with an RMS scatter in $\log \rm{Age}$ of 0.238 dex. We give the Pearson's and Spearman's Rank Correlation Coefficients in the top of panel c).
}
\label{fig:vs_eps_model}
\end{center}
\end{figure*}

It is important to emphasize that this newly discovered relation extends beyond the general notion where "disks are young" and "bulges are old". First, \vse\ is the more accurate, physically motivated and dynamical equivalent of the morphological bulge-to-disk ratio measurements that are notoriously difficult\cite{lange2016} and hampered by inclination and dust extinction. Secondly, bulge-to-disk decompositions have revealed a wide distribution of the S\'ersic index for both bulges \emph{and} disks, but more importantly, have demonstrated that bulges can have blue colours while disks can be red, and that decomposed early-type bulges and disks have the same colour distribution\cite{lange2016}. Thirdly, optical morphology is not an ideal tracer of the underlying dynamical structure for early-type galaxies\cite{emsellem2011}, as a large fraction of galaxies with elliptical morphology have high \vse\ values \cite{cappellari2016}. Fourthly, we recover the same relation for low mass galaxies ($\log M_{*}/\msun<10.25$, Supplementary Figure \ref{fig:sup_vs_eps_age_massbins}), where the fraction of classical, dispersion-dominated bulges is less than 30\%\cite{fisher2011}.

To remove any doubt that the observed relation between age and intrinsic ellipticity is due to our large range in optical morphology, we now separate our sample into early-type (Figure \ref{fig:vs_eps_age_es}a) and late-type galaxies (Figure \ref{fig:vs_eps_age_es}b). Our results confirm the well-known fact that early-type galaxies are on average older, more spherical, and more dispersion-dominated than late-type galaxies. Within the early-type and late-type morphological samples, the LOESS smoothed trends are consistent with the trend of the full sample: young stellar populations reside in rapidly rotating intrinsically-flat galaxies, and older stellar populations are found in more spherical and pressure-supported galaxies. The trend between age and intrinsic shape also remains if we only select S0s and early-spirals.

To quantify this relation further, in Figure \ref{fig:vs_eps_model}, we deproject our observed (\vs, \ee) data onto the intrinsic ellipticity-inclination plane, using the theoretical model predictions for rotating, oblate, axisymmetric spheroids with varying intrinsic shape and anisotropy. Galaxies below the magenta line are excluded from the sample because they are outside the model range, which excludes the majority of slow-rotating galaxies\cite{emsellem2011}, that are typically massive with old stellar populations (see Figure \ref{fig:vs_lmass_age}). Furthermore, we exclude near-perfectly round galaxies ($\ee<0.025$) for which the model predictions are highly degenerate and the relative measurement uncertainties on \ee\ are large. Our deprojection approach is supported by an independent method for determining the intrinsic ellipticity from inverting the distributions of apparent ellipticities and kinematic misalignments that show that the intrinsic ellipticity agrees well with the theoretical model prediction\cite{foster2017}. 

After deprojecting the data, we still find that stellar age follows the intrinsic ellipticity remarkably well, with no effect on inclination. For low inclination the LOESS smoothing suggests a slight deviation in the age-intrinsic ellipticity relation. However, for these galaxies, that are observed nearly face-on, the deprojection is more uncertain because the model tracks in the (\vs, \ee) plane are closest together. We note that the relation between anisotropy and intrinsic ellipticity is a key ingredient in the deprojection. However, the relation between age and intrinsic ellipticity does not change within the anisotropic uncertainty, and only in the extreme case of no anisotropy ($\beta_z = 0$) do we start to see a mild deviation of the horizontal age trend in Figure \ref{fig:vs_eps_model}. Axisymmetric Jeans anisotropic modelling, or more flexible Schwarzschild modelling are required to simultaneously constrain the intrinsic ellipticity and anisotropy, but this is beyond the scope of this Letter.

We show the relation between age and intrinsic ellipticity in Figure \ref{fig:vs_eps_model}c. The contour reveals a curved relation, yet the root-mean-square (RMS) scatter in $\log \rm{Age}$ from a best-fitting linear and quadratic function are the same (0.238 dex). Early-type and late-type galaxies reside at opposite sides of the relation, but in the region of overlap show consistent age and intrinsic ellipticity values. 
When using different aperture sizes (0.5, 1, and 2 \re), we find consistent results, which indicates that the relation between age and intrinsic ellipticity is not influenced by aperture effects.

Due to our broad range in stellar mass, optical morphology, and large-scale environment, our results provide strong constraints on how galaxies build-up their stellar mass and angular momentum. 
The formation and growth of a bulge due to mergers will lower \vse\ and \ee, but increase the mean intrinsic ellipticity as a function of time\cite{kormendy2004,cappellari2016}. Secular evolutionary processes, such as bar formation through dynamical instabilities, also change the observed \vse\ and ellipticity distribution of galaxies over time\cite{kormendy2004}; the stellar velocity dispersion is observed to increase with bar strength\cite{seidel2015}. Thus, both mergers and secular evolution can explain, at least partially, the relation between age and intrinsic ellipticity observed in the SAMI Galaxy Survey data.

For early-type galaxies, the relation between age and intrinsic ellipticity is consistent with predictions from hydrodynamical cosmological simulations and observations at $z\sim2-3$ when the Universe was only 3 billion years old. Massive ($\log M_{*}/\msun>11$) quiescent galaxies at high-redshift have significantly higher effective stellar velocity dispersions and effective densities as compared to similar present-day galaxies\cite{vandokkum2015}. Due to the surprising compactness of these galaxies in the early-Universe\cite{vanderwel2014}, obtaining stellar rotation measurements has only been possible for extremely rare, strongly gravitationally-lensed galaxies\cite{newman2015}. The handful of stellar rotation curve measurements at these redshift suggest that massive red-and-dead galaxies at $z\sim2-3$ are still rapidly rotating. Both major and minor mergers are predicted to make these galaxies intrinsically more spherical and dynamically more pressure-supported\cite{naab2014}. While dry minor mergers are required to explain the strong evolution in size, dry major mergers are most effective at creating slow-rotating dispersion-dominated galaxies\cite{penoyre2017}. This picture is consistent with the relation between age, \vse\, and intrinsic ellipticity that we find in the SAMI Galaxy Survey early-type population: older stellar populations are more likely to reside in massive, more dispersion-dominated and more spherical systems, that can only form through dramatic interactions which are more common in the early-Universe.

In the Milky Way, recent studies suggest that the Galactic thick disk is distinct from the dominant thin disk in abundance ratios and age\cite{blandhawthorn2016}. The younger, thin-disk stars are on near-circular, co-rotational orbits with a low velocity dispersion\cite{lee2011}, whereas the thick disk with higher dispersion\cite{sharma2014} is suggested to arise from a combination of stellar migration and/or flaring of the old disk stars\cite{schonrich2009}. Some doubt remains as to whether the thick and thin disk are separate entities or a continuous distribution, but the parallels to the work presented here are striking. Within the SAMI sample, we find a similar trend where older stellar populations are in intrinsically thicker structures.

These present-day, dynamically-hot galaxies could also be descendants of high-redshift disks that exhibit large random motions\cite{forsterschreiber2009} and are geometrically thick\cite{tacconi2013}. For rotation-dominated disks, the observed velocity dispersion of the ionised H$\alpha$ gas decreases by a factor of two over a period of 3.5 Gyr from redshift $z\sim2.3$ to $z\sim0.9$\cite{wisnioski2015}. As H$\alpha$ is a direct tracer of star formation, this suggests that as the Universe ages, newly formed stellar populations are more likely to reside in colder rotating disks, conforming with our findings here.

In this Letter we have revealed a relation between the intrinsic shape and stellar population age of galaxies, by combining imaging, spatially resolved dynamics, and stellar population measurements. This shows the power of utilising integral field spectroscopy on a large sample of galaxies to further our understanding of physical processes involved in the build-up of stellar mass and angular momentum in galaxies.


\bibliography{jvds_sami_vsp_arxiv}


\begin{addendum}
 \item[Acknowledgements] The SAMI Galaxy Survey is based on observations made at the Anglo-Australian Telescope. The Sydney-AAO Multi-object Integral field spectrograph (SAMI) was developed jointly by the University of Sydney and the Australian Astronomical Observatory. The SAMI input catalogue is based on data taken from the Sloan Digital Sky Survey, the GAMA Survey and the VST ATLAS Survey. The SAMI Galaxy Survey is funded by the Australian Research Council Centre of Excellence for All-sky Astrophysics (CAASTRO), through project number CE110001020, and other participating institutions. The SAMI Galaxy Survey website is http://sami-survey.org/. Parts of this research were conducted by the Australian Research Council Centre of Excellence for All Sky Astrophysics in 3 Dimensions (ASTRO 3D), through project number CE170100013. J.v.d.S. is funded under Bland-Hawthorn's ARC Laureate Fellowship (FL140100278). N.S. acknowledges support of a University of Sydney Postdoctoral Research Fellowship. S.B. acknowledges the funding support from the Australian Research Council through a Future Fellowship (FT140101166). M.S.O. acknowledges the funding support from the Australian Research Council through a Future Fellowship (FT140100255). J.v.d.S. and N.S. would like to thank all SAMI team members for valuable discussions. We thank the anonymous referees for their constructive comments. 
 \item[Correspondence] Correspondence and requests for materials should be addressed to J.v.d.S. ~(email: jesse.vandesande@sydney.edu.au).
\item[Author Contributions] J.v.d.S. \& N.S. led the interpretation. J.v.d.S. measured the stellar kinematic parameters from the SAMI Galaxy Survey spectra, and wrote the text. N.S. measured the Lick indices from the spectra, and derived the stellar population ages. F.E. measured the structural parameters. All authors contributed to the analysis and interpretation of the Letter, and contributed to overall Team operations including target catalogue and observing preparation, instrument maintenance, observing at the telescope, writing data reduction and analysis software, managing various pieces of team infrastructure such as the website and data storage systems, and innumerable other tasks critical to the preparation and presentation of a large data set presented here.
\end{addendum}

\clearpage
\newpage

\begin{center}
{\bf \Large \uppercase{Methods} }
\end{center}

\textbf{The SAMI Galaxy Survey.} The Sydney-AAO Multi-object Integral field spectrograph (SAMI\cite{croom2012}) is mounted at the prime focus on the Anglo-Australian Telescope. SAMI uses 13 fused-fibre bundles (Hexabundles\cite{blandhawthorn2011, bryant2014}) with a high (75\%) fill factor, deployable over a 1 degree diameter field of view. Each hexabundle contains 61 fibres of 1\farcs6 angle on sky, resulting in each IFU covering a $\sim15^{\prime\prime}$ diameter region on the sky. The IFUs, as well as 26 individual sky fibres, are plugged into pre-drilled plates using magnetic connectors. SAMI fibres are fed to the double-beam AAOmega spectrograph\cite{sharp2006}. AAOmega allows a range of different resolutions and wavelength ranges. For the SAMI Galaxy Survey, the 580V and 1000R grating are used in the blue (3750-5750\AA) and red (6300-7400\AA) arm of the spectrograph, respectively. This results in a resolution of R$_{\rm{blue}}\sim 1810$ ($\sigma\sim70\kms$) at 4800\AA, and R$_{\rm{red}}\sim4260$ ($\sigma\sim30\kms$) at 6850\AA\cite{vandesande2017}.

The SAMI Galaxy Survey\cite{croom2012,bryant2015} aims to observe 3600 galaxies, covering a broad range in galaxy stellar mass (M$_* = 10^{8}-10^{12}$\msun) and galaxy environment (field, groups, and clusters). Field and group targets were selected from four volume-limited galaxy samples derived from cuts in stellar mass in the Galaxy and Mass Assembly (GAMA\cite{driver2011}) G09, G12 and G15 regions. Cluster targets were obtained from eight high-density cluster regions sampled within radius $R_{200}$ with the same stellar mass limit as for the GAMA fields \cite{owers2017}. Observations of the data presented here were carried out between Jan-2013 and Jun-2015. Reduced data are presented as three dimensional data cubes (two spatial, one spectral dimension) with a spatial sampling of 0\farcs5\cite{sharp2015}. The data have a median seeing of 2\farcs1\cite{allen2015}.

\textbf{Ancillary Data.} We use the Multi-Gaussian Expansion (MGE\cite{emsellem1994,cappellari2002,scott2009}) technique to derive effective radii, ellipticities, and position angles, using imaging data from GAMA-SDSS\cite{driver2011}, SDSS\cite{york2000}, and VST\cite{shanks2013,owers2017}. \re\ is defined as the semi-major axis effective radius, and the ellipticity of the galaxy within one effective radius as $\epsilon_{\rm{e}}$, measured from the best-fitting MGE model. Visual morphological classifications were performed on the SDSS and VST colour images; late- and early-types are divided according to their shape, presence of spiral arms and/or signs of star formation\cite{cortese2016}. Stellar mass estimates $M_*$ are obtained from the rest-frame $i$-band absolute magnitude and $g-i$ colour by using a colour-mass relation \cite{taylor2011,bryant2015}. A Chabrier stellar initial mass function\cite{chabrier2003} and exponentially declining star formation histories are assumed.

\textbf{Stellar Kinematics.} We measure the stellar kinematic parameters from the SAMI data\cite{vandesande2017} by fitting the spectra with the penalized pixel fitting code (pPXF\cite{cappellari2004}), assuming a Gaussian line of sight velocity distribution (LOSVD), i.e., extracting only the stellar velocity $V$ and stellar velocity dispersion $\sigma$. The SAMI blue and red spectra (convolved to a common spectral resolution) are rebinned onto a logarithmic wavelength scale with constant velocity spacing (57.9 \kms), using the code \textsc{log\_rebin} provided with the \textsc{pPXF} package. Annular binned spectra are used to derive local optimal templates. For building templates we use the MILES stellar library \cite{sanchezblazquez2006}, which consists of 985 stars with large variety in stellar atmospheric parameters.

After constructing annular optimal templates, we run \textsc{pPXF} three times on each galaxy spaxel. The first fit is used for determining a precise measure of the noise scaling from the residual of the fit. A 12th order additive Legendre polynomial is used to remove residuals from small errors in the flux calibration. In the second fit, we mask the emission lines and clip outliers using the CLEAN parameter in \textsc{pPXF}. The velocity and velocity dispersion are extracted from the third and final fit. In the last step, \textsc{pPXF} is fed with the optimal templates from the annular bin in which the spaxel is located, plus the optimal templates from adjacent annular bins. Uncertainties on the LOSVD parameters are estimated from 150 simulated spectra. For each spaxel, we estimate the uncertainties on the LOSVD parameters from the residuals of the best-fit minus the observed spectrum. These residuals are then randomly rearranged in wavelength and added to the best-fit template. We refit this simulated spectrum with \textsc{pPXF}, and the process is then repeated 150 times. The uncertainties on the LOSVD parameters are the standard deviations of the resulting simulated distributions.

From the unbinned flux, velocity, and velocity dispersion maps, we derive \vse\ within an elliptical aperture with semi-major axis \re\ and axis ratio $b/a$, using the following definition\cite{cappellari2007}:
\begin{flushleft}
\begin{equation}
\left(\frac{V}{\sigma}\right)^2 \equiv \frac{\langle V^2 \rangle}{\langle \sigma^2 \rangle} = \frac{ \sum_{i=0}^{N_{spx}} F_{i}V_{i}^2}{ \sum_{i=0}^{N_{spx}} F_{i}\sigma_i^2},
\label{eq:vs}
\end{equation}
\end{flushleft}
The subscript $i$ refers to the spaxel position within the ellipse, $F_{i}$ the flux of the $i^{th}$ spaxel, $V_{i}$ is the stellar velocity in \kms, $\sigma_{i}$ the velocity dispersion in \kms. We sum over all spaxels $N_{spx}$ that meet the following quality cut\cite{vandesande2017}: signal-to-noise (S/N) $>3$\AA$^{-1}$, \sobs $>$ FWHM$_{\rm{instr}}/2 \sim 35$\kms\ where the FWHM is the instrumental spectral full-width at half-maximum, $V_{\rm{error}}<30$\kms, and $\sigma_{\rm{error}} < (\sobs *0.1 + 25$\kms). For galaxies where the largest aperture radius measurement is less than one \re\ ($\sim15$ percent of the total sample), the \vse\ measurements are aperture corrected\cite{vandesande2017b} to one \re.

\textbf{Stellar Populations}. 
Luminosity-weighted stellar population ages are estimated using 11 Lick indices in the SAMI blue spectral range\cite{scott2017}. For each aperture spectrum, we first construct an emission corrected spectrum using a similar approach as the stellar kinematic fits. In series of three pPXF fits, a smoothed noise spectrum is created which is then used to identify regions of bad pixels or emission lines using the CLEAN keyword in pPXF. These flagged pixels are replaced with corresponding pixels in the best-fitting template spectrum. The emission corrected observed spectrum is then shifted to a rest-frame wavelength scale based on the measured velocity. Due to the different spectral resolution of each Lick index, we convolve the observed spectrum with a Gaussian, such that the total broadening of the spectrum matches that of the Lick system. In practice, $\sigma^2_{\rm{instrument}}+\sigma^2_{\rm{galaxy}}+\sigma^2_{\rm{applied}} = \sigma^2_{\rm{Lick}}$, where $\sigma_{\rm{instrument}}$ is the instrumental broadening, $\sigma_{\rm{galaxy}}$ is the velocity broadening due to each galaxies intrinsic velocity dispersion, $\sigma_{\rm{applied}}$ is the additional broadening applied to each spectrum to match the Lick resolution and $\sigma_{\rm{Lick}}$ is the broadening of each index in the Lick system. Uncertainties on all indices are estimated in a similar fashion as the stellar kinematic uncertainties using a 100 simulated spectra using a Monte Carlo procedure.

Lick indices are converted into single stellar population equivalent age using stellar population synthesis models\cite{schiavon2007} that predict Lick indices as a function of $\log \rm{Age}$, metallicity [Z/H], and [$\alpha$/Fe]. For each aperture spectrum, we use a $\chi^2$ minimisation approach to determine the SSP that best reproduces the measured Lick indices. In the three-dimensional space of age, [Z/H], and [$\alpha$/Fe], we adopt the values that result in the minimum $\chi^2$, and uncertainties on the three SSP parameters are determined from the $\chi^2$ distribution. Typical uncertainties are $\pm0.15$ in $\log$ age, $\pm0.17$ in [Z/H] and $\sim0.14$ in [$\alpha$/Fe].

We caution that our luminosity-weighted SSP-equivalent ages are likely to be biased to young ages (relative to mass-weighted values) for all galaxies that have experienced recent star formation, as we do not account for differing star formation histories, beyond varying [$\alpha$/Fe]. However, the relative differences in SSP-equivalent ages between galaxies are robust and reflect real differences in stellar populations. For early-type galaxies, on average, the SSP derived ages are lower by 0.2 dex in log Age [Gyr] than mass-weighted ages \cite{mcdermid2015}. However, as the offset between luminosity-weighted and mass-weighted age follows a linear relation with little scatter, this implies that our conclusion will not change if mass-weighted ages are used.

\textbf{Code availability.} The data reduction package used to process the SAMI data is available at http://ascl.net/1407.006, and makes use of 2dfdr: http://www.aao.gov.au/science/software/2dfdr. To derive the stellar kinematic parameters and the lick absorption line strengths, we use the publicly available penalised pixel-fitting (pPXF) code from M. Capppellari: {http://www-astro.physics.ox.ac.uk/~mxc/software/\#ppxf}. For the adaptive LOESS smoothing, we use the code from M. Cappellari obtained from: http://www-astro.physics.ox.ac.uk/~mxc/software/\#loess

\textbf{Data availability.} All reduced data-cubes in the GAMA fields used in this Letter are available on: http://datacentral.aao.gov.au/asvo/surveys/sami/, as part of the first SAMI Galaxy Survey data release \cite{green2017}. Stellar kinematic data products will become available in the second SAMI Galaxy Survey data release. 


\clearpage
\newpage

\onecolumn

\begin{center}
{\bf \Large \uppercase{Supplementary information} }
\end{center}

\setcounter{figure}{0}
\vspace{2cm}

\begin{figure*}[!h]
\begin{center}
\includegraphics[width=0.8\linewidth]{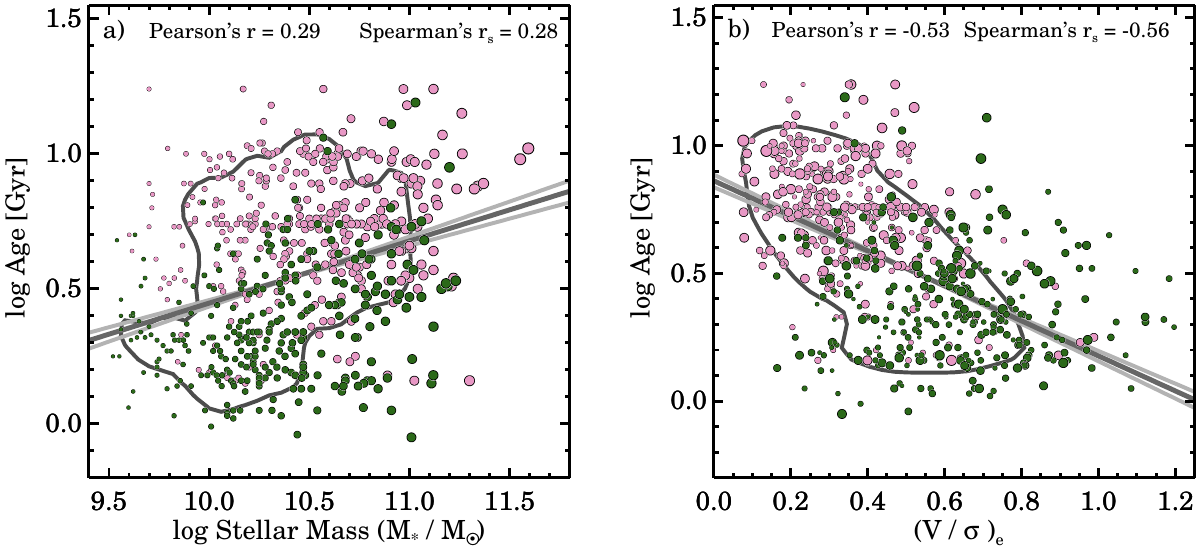}
\caption{Relation between age and stellar mass in panel a) and age and \vse\ in panel b). Sample selection is the same as in Figure \ref{fig:vs_eps_model}c in the main text. Early-type galaxies are shown in pink, late-type galaxies in green. The contour encloses 68\% of the total probability using kernel density estimates. We give the Pearson's and Spearman's Rank Correlation Coefficients in the top of panels a) and b). In panel a), the solid line shows the best-fitting relation of $\log \rm{Age} = 0.23\pm0.03\times\log_{10}{M_*} - 1.84\pm0.31$, with an RMS scatter in $\log \rm{Age}$ of 0.277 dex, using the \textit{Interactive Data Language} \textsc{LINMIX\_ERR} function that uses Bayesian approach to linear regression. In panel b), the solid line shows the best-fitting relation of $\log \rm{Age} = -0.69\pm{0.05}\times\vse + 0.86\pm0.02$, with an RMS scatter in $\log \rm{Age}$ of 0.245 dex. Note that the age-intrinsic ellipticity relation has less scatter (RMS = 0.238 dex) than the age-stellar mass and age-\vse\ relations. Given that the uncertainty on intrinsic ellipticity is larger than on \vse, the intrinsic scatter in the age intrinsic ellipticity relation is smaller than in the age-\vse\ relation, despite the relatively small different in observed RMS scatter.
}
\label{fig:sup_age_mass_vs_scatter}
\end{center}
\end{figure*}

\begin{figure*}[!h]
\begin{center}
\includegraphics[width=\linewidth]{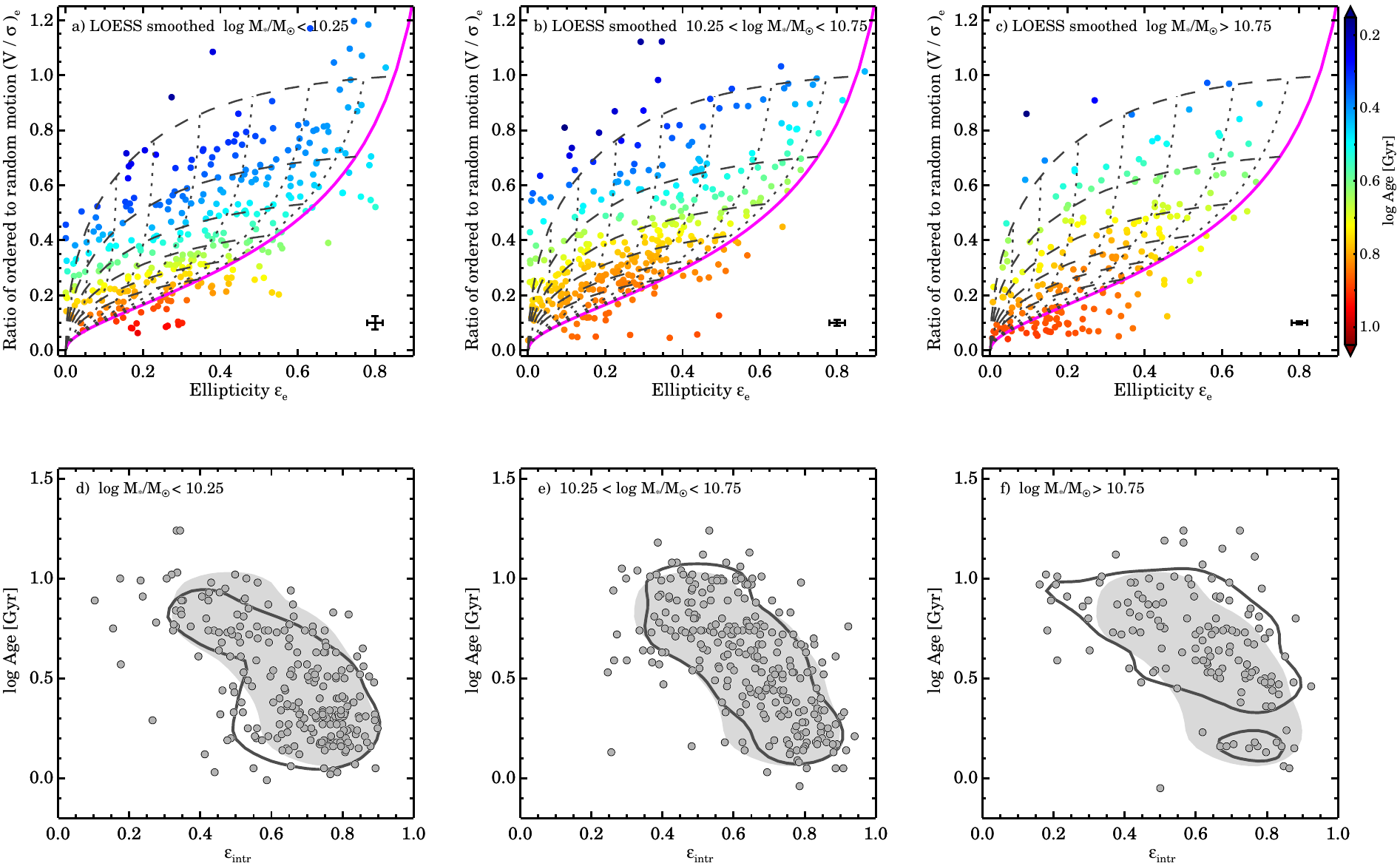}
\caption{Intrinsic ellipticity and inclination derived from theoretical predictions and our data from within the (\vs, \ee) diagram, similar to the main text Figure \ref{fig:vs_eps_age} (top row), and the main text Figure \ref{fig:vs_eps_model}c (bottom row). Different panels show three different stellar mass bins, as indicated in each panel. The colour coding highlights the LOESS smoothed stellar population ages. The LOESS algorithm is applied separately to the three individual subsamples. The median uncertainty on \vse\ and \ee\ is shown in the bottom-right corner, and the median uncertainty on $\log$ age is $\pm$0.15 dex. Lines in the top row are similar to Figure \ref{fig:vs_eps_age} in the main text. Contours in the bottom row enclose 68\% of the total probability using kernel density estimates. We find that the relation between age and intrinsic ellipticity is independent of stellar mass. Furthermore, for each mass bin we find a relation between age and instrinsic ellipticity consistent with the relation for the full sample (grey filled region). The high-mass bin has a larger fraction of old galaxies, but the distribution of points is consistent with the full sample.
}
\label{fig:sup_vs_eps_age_massbins}
\end{center}
\end{figure*}

\begin{figure*}[!h]
\begin{center}
\includegraphics[width=0.8\linewidth]{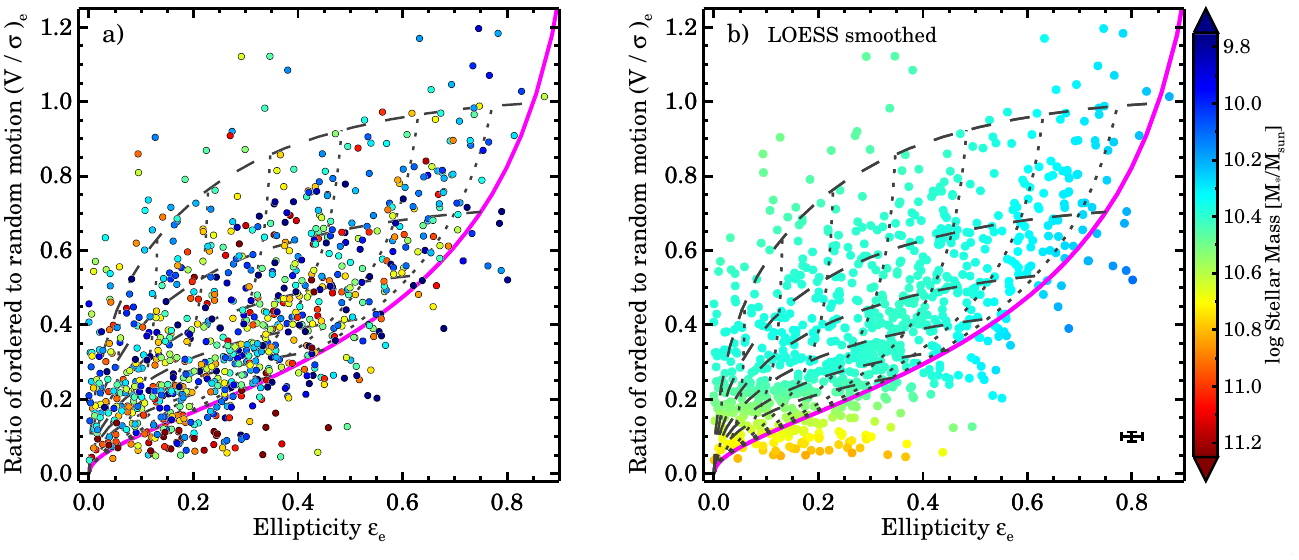} 
\caption{Linking stellar dynamics (\vs) and observed shape (ellipticity $\ee$) with total stellar mass. In panel a) the colour coding reflects low stellar mass (blue) and high stellar mass (red), whereas in panel b) we use the LOESS smoothing algorithm to look for a mean underlying trend in stellar mass.
The median uncertainty on \vse\ and \ee\ is shown in the bottom-right corner, and the median uncertainty on $\log$ stellar mass is $\pm$0.1 dex. Lines similar to Figure \ref{fig:vs_eps_age} in the main text.
We find no trend with stellar mass for galaxies above magenta line, i.e., objects that are consistent with being axisymmetric rotating oblate spheroids. However, galaxies with the highest stellar mass are more likely to be pressure supported spheroids than rotating.
}
\label{fig:sup_vs_eps_mass}
\end{center}
\end{figure*}

\begin{figure*}[!t]
\begin{center}
\includegraphics[width=\linewidth]{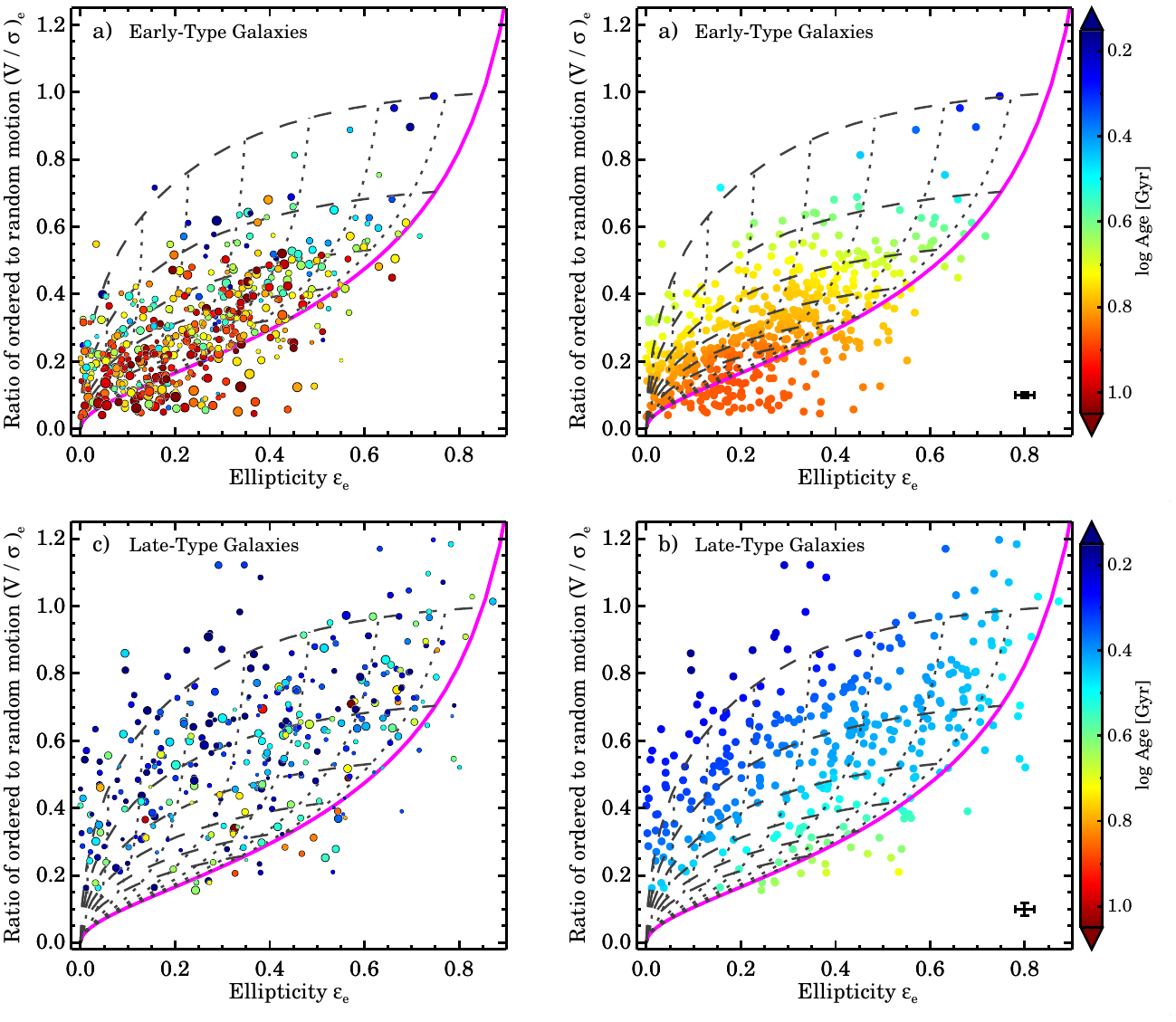} 
\caption{Luminosity-weighted stellar age in the (\vs, \ee) diagram, split by visual morphological type, similar to Figure \ref{fig:vs_eps_age_es} but now including stellar population age of each individual galaxy on the left, and the LOESS smoothed stellar population ages on the right. Galaxies with early-type morphology (Ellipticals and S0s) are shown in the top row, galaxies with late-type morphology (Sa-Sd, irregulars) are shown bottom row. The median uncertainty on \vse\ and \ee\ is shown in the bottom-right corner, and the median uncertainty on $\log$ age is $\pm$0.15 dex. Lines similar as in Figure \ref{fig:vs_eps_age} in the main text. The LOESS algorithm is applied separately to the two individual subsamples.
}
\label{fig:sup_vs_eps_age_es_orig}
\end{center}
\end{figure*}

\begin{figure*}[!t]
\begin{center}
\includegraphics[width=0.55\linewidth]{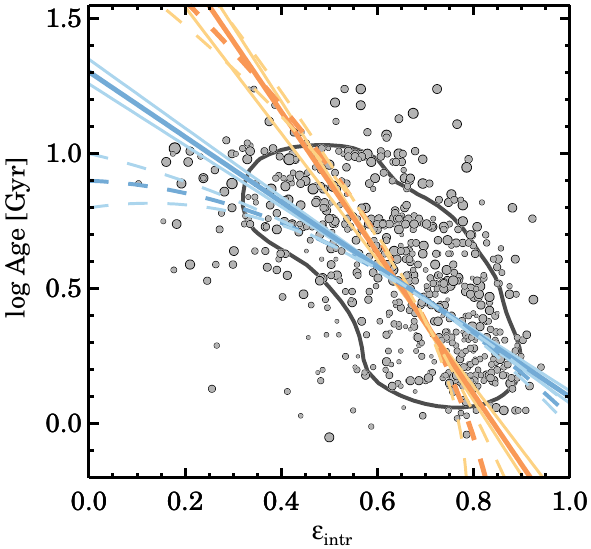} 
\caption{Age versus intrinsic ellipticity. The contour encloses 68\% of the total probability using kernel density estimates. The blue solid and dashed lines shows the best-fitting relation as described in Figure \ref{fig:vs_eps_model}c in the main text, where we minimised the scatter in log Age (vertical). If we minimise the scatter in intrinsic ellipticity (horizontal), then we obtain: $\epsilon_{\rm{intr}} = -0.37\pm0.03\times\log \rm{Age} + 0.84\pm0.01$ (solid orange line), with an RMS scatter in $\epsilon_{\rm{intr}}$ of 0.138 , and $\epsilon_{\rm{intr}} = -0.12\pm0.07\times{\log \rm{Age}}^2 - 0.19\pm0.08\times\log \rm{Age} + 0.79\pm0.02$ (dashed orange line) with an RMS scatter in $\epsilon_{\rm{intr}}$ of 0.136. 
}
\label{fig:sup_age_es_linefits}
\end{center}
\end{figure*}

\end{document}